Micellar effects on Ostwald ripening in emulsions: Transition from cubic to quadratic particle size growth

Alexey Kabalnov, Ink Splat Company, kabalnov@inkjet3d.com

Abstract

Ostwald ripening in O/W emulsions in presence of solubilizing micelles is theoretically studied. At small average sizes, the kinetics is predicted to follow the classical Lifshits-Slezov-Wagner cubic law, with the rate proportional to the molecular solubility of the oil in water, as if no micelles were present. At larger particle sizes the kinetics transitions to the Wagner's quadratic law. The crossover point for the kinetics depends on the dynamics of the oil solubilizate-micelle exchange; it is set by the value of the oil atmosphere distribution parameter, $\kappa$, which, somewhat like Debye length, is proportional to the square root of the micellar concentration. It should be noted that in the range when $1/\kappa$ is close to the particle average radius, the ripening kinetics still nearly follows the cubic law, with only moderate deviations; in this case, the micellar effects are experimentally seen not as the deviations from the linearity, but as an apparent increase in the cubic rate. The increase is predicted to be larger for nonionic micelles of ethylene oxide type. The square root growth with the micellar concentration is limited by the range where no micellar growth/transition to wormlike shape with the increase in the surfactant concentration occurs.

**Introduction**

Ostwald ripening is the process of a gradual dissolution of smaller particles and the growth of larger ones at their expense [1- 3]. The cornerstone of the ripening kinetics is the rate of dissolution for a single particle. If the rate is controlled by the diffusion in the bulk, the steady state rate in the spherical symmetry can be written as:

$$J = \frac{4\pi}{3}\frac{d}{dt}(r^3) = 4\pi r^2 D\left(C_m - C_p\right)\frac{1}{r} = 4\pi r D\Delta C \qquad (1)$$

Here $J$ is the mass flow, reduced to the density of the diffusing material, with the dimension of $cm^3/s$, $r$ is the radius of the particle, $C_m$ is the concentration in the medium, $C_p$ is the concentration at the surface of the particle, $D$ is the diffusion coefficient. The concentrations are reduced to the density of the diffusing material and are dimensionless; this convention for concentrations is followed in this whole paper. The $1/r$ factor can be interpreted as the reciprocal of the effective diffusion path, which is the particle radius for the case of the spherical symmetry. The concentration at the surface of the particle is determined by Kelvin equation:

$$C_p = C_\infty exp\left(\frac{2\sigma V_m}{rRT}\right) \qquad (2)$$

Here $C_\infty$ is the solubility of the macrophase at $r=\infty$, $\sigma$ is the interfacial tension, $V_m$ is the molar volume of the dispersed phase, $R$ is the gas constant, and $T$ is the absolute temperature. One can introduce a characteristic length to the problem, $\alpha = \frac{2\sigma V_m}{RT}$, which, for common emulsions and dispersions, has the order of few nanometers. As in most practical cases, the particle size is larger than few nm, $\alpha \ll r$, and it is possible to expand the exponent in series:

$$C_p = C_\infty exp\left(\frac{\alpha}{r}\right) \approx C_\infty\left(1+\frac{\alpha}{r}\right) \qquad (3)$$

The equation (1) determines the growth/dissolution rate for the case when the diffusion is delimited by the molecular diffusion in the bulk. However, other kinetic regimes are also possible. For example, as suggested by Wagner [2], if the dissolution /growth of a crystal is explored, the rate can be controlled by a 'surface reaction' at the face of the crystal; the 'rate constant' $k_r$, has the dimension of cm/s. Knowing that this step is delimiting, we can write down the particle growth rate as

$$J = \frac{4\pi}{3}\frac{d}{dt}(r^3) = 4\pi r^2 k_r\left(C_m - C_p\right) = 4\pi r^2 k_r \Delta C \qquad (4)$$

Note that the crystal growth is just one case when this quadratic-in-radius law can develop. For example, the same kinetics occurs if the particle is surrounded by a poorly penetrable membrane; then the same equation will hold, but $k_r$ becomes the diffusivity through the membrane divided by its thickness. This scenario is in principle possible for emulsion systems, with emulsion drops covered with a poorly penetrable membrane.

In polydisperse systems undergoing ripening, Eqns (1) and (4) determine the velocities of the particles in the space of sizes, and eventually, the kinetics of the change in the average particle size and evolution of the particle size distribution. The theory predicts that an asymptotic regime develops in both cases, where certain limiting laws apply. In the case of eqn (1) law, the *cube* of number-average radius linearly increases with time, with the rate:

$$w = \frac{d(\bar{r}^3)}{dt} = \frac{8C_0\sigma V_m D}{9RT} \qquad (5)$$

Here $\bar{r}$ is the number average radius. On the other hand, in case when the growth rate is determined by eqn (4), the *square* of the average radius increases linearly with time:

$$v = \frac{d(\bar{r}^2)}{dt} = \frac{64C_0\sigma V_m k_r}{81RT} \qquad (6)$$

The theory also predicts the evolution of the particle size distributions; in both cases, the distribution functions of the reduced average radius $u = r/\bar{r}$ become time invariants tending to specific functions called attractors. The attractors are different for the cubic and quadratic laws of growth [1- 3].

There could be scenarios when the cubic and quadratic laws coexist in the same physical system, with one transitioning to another. For example, in case of ripening in crystalline dispersions, at smaller particle sizes, the growth can be delimited by the rate of the molecular attachment, whereas for larger sizes the kinetics switches to the diffusion control. This discussion is convenient to have in terms of the diffusional resistances for each stage. Thus, the diffusional resistances of the attachment and external diffusion stages can be expressed as $r/D$ and $1/k_r$, respectively; the diffusional resistance of the whole system can be considered as the two resistances connected in series, or their sum. At the small particle size, $r/D < 1/k_r$ and the attachment stage is the rate delimiting step. Conversely, after the average particle size grows such that $r/D > 1/k_r$, the diffusion in the bulk becomes the rate delimiting step. Accordingly, the rate of the average particle size growth is expected to transition from the quadratic to cubic law with time as the particle size increases, and the distribution function to undergo an adjustment from one attractor to another. Note also that in a continuously distributed polydisperse system undergoing ripening, there will always be a fraction of particles for which $r/D < 1/k_r$, even when for the number average radius $\bar{r}$, $\bar{r}/D \gg 1/k_r$. We just note here that the fraction of these particles will progressively decrease as the average particle size continues to increase.
In this paper, we will be considering another case, where both quadratic and cubic laws can co-exist in the same system, with the transition from one mode to another, although in a somewhat different way. In aqueous micellar systems, micelles are capable to absorb oils in the interior, in the process known as solubilization. The rate of this process is slow, as in many cases the oils cannot be absorbed by micelles directly, but only through the stage of molecular solution in water. In our previous paper [4], we analyzed coupling between the dissolution kinetics of a spherical oil drop in a micellar solution and the micellar dynamics, where the micelles are constantly exchanging the oil with the medium. The equation below shows the predicted dissolution rate:

$$J = \frac{4\pi}{3}\frac{d}{dt}(r^3) = 4\pi r^2 D \Delta C\left(\kappa + \frac{1}{r}\right) \qquad (7)$$

Here $\kappa$ is the parameter having the dimension of reciprocal length, which reflects on the oil molecular solution concentration profile around the dissolving particle. The value of $\kappa$ depends on the rate of molecular exchange between the oil in the aqueous solution and the micellar interior:

$$\kappa = \sqrt{\frac{\omega n_{mic}}{D}} \qquad (8)$$

here $\omega$ is the source/sink term for the oil generation by a micelle [$cm^3/s$], $n_{mic}$ is the number concentration of micelles per unit volume, $cm^{-3}$, and $D$ is the oil-in water molecular diffusion coefficient. An explanation of these equations, and estimates for the source-sink term $\omega$ will follow; at this moment we just note that the particle growth rate of Eqn (7) is a cross between the two rates shown in eqns (1) and (4). Thus, for small average particle sizes, $\bar{r} \ll 1/\kappa$, the kinetic of dissolution will follow eqns (1) and (5), and ripening should follow the classical cubic law; conversely, if $\bar{r} \gg 1/\kappa$, the kinetics should switch to quadratic law per eqns (4) and (6), and the rate will be equal to:

$$v = \frac{d(\bar{r}^2)}{dt} = \frac{64 C_0 \sigma D V_m \kappa}{81 RT} \qquad (9)$$

Equation 9 can be arrived to by direct substitution of $k_r$ in Eqn 6 by $D\kappa$, per eqn 7, in the limit of $r \gg 1/\kappa$. If the ripening kinetics is observed over a long time and over wide range of sizes, it can switch from cubic to quadratic with some transition regime in between. The case therefore is somewhat like ripening in crystalline dispersions considered above. The main difference is that the diffusion resistances in the case of interest are connected, so-to-say, not in series, but in parallel, and the kinetics transitions from the cubic law at small sizes to quadratic law at larger sizes, in contrary to what is predicted for the crystalline growth (Figure 1).

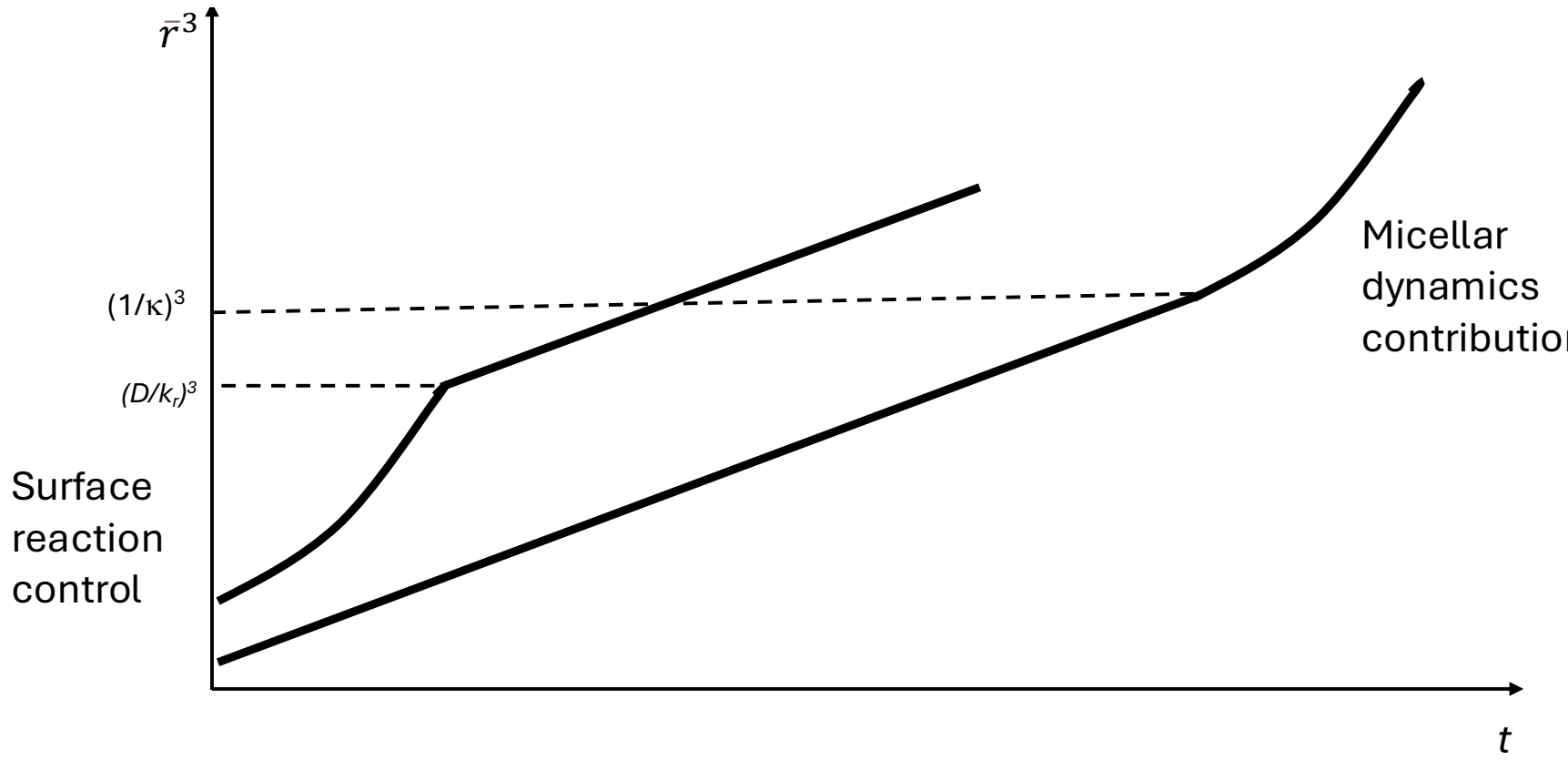

Figure 1. Cartoon explaining the difference between the kinetic transitions from quadratic to cubic growth for the surface reaction control, and from cubic to quadratic growth in case of the micellar dynamics' contribution. In cubic coordinates, the quadratic part of the kinetics, which looks concave, either happen at the beginning, or in the end of the kinetic plot.

There is a relatively large volume of experimental data on Ostwald ripening in emulsions [5-17]. Most of experimental data on the particle size growth were analyzed by plotting the data in cubic coordinates. If the range of particle size data is relatively narrow, a good fit can be obtained in cubic coordinates; we note that the micellar effects in this case will be seen not so much as the deviations of the plot from linearity, but as an increase in the cubic rate with the micellar concentration, as illustrated in Figure 2 A and B. In case when ripening kinetics is studied over larger particle size range where $r \gg 1/\kappa$, the initial linear part may not be visible at all,  and the graphs will have a uniform curvature, compatible with quadratic growth, see Figure 2C.

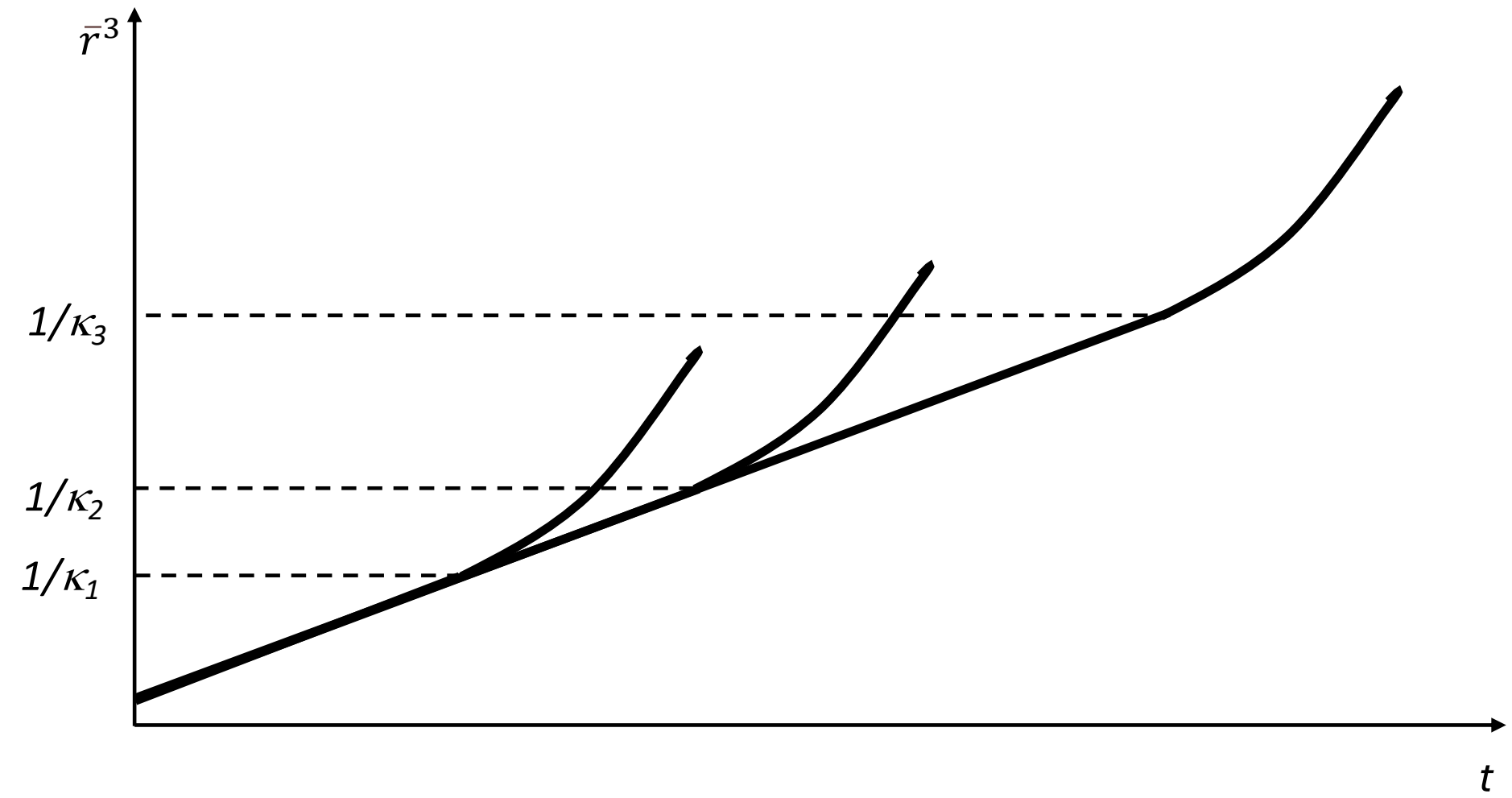

(A)

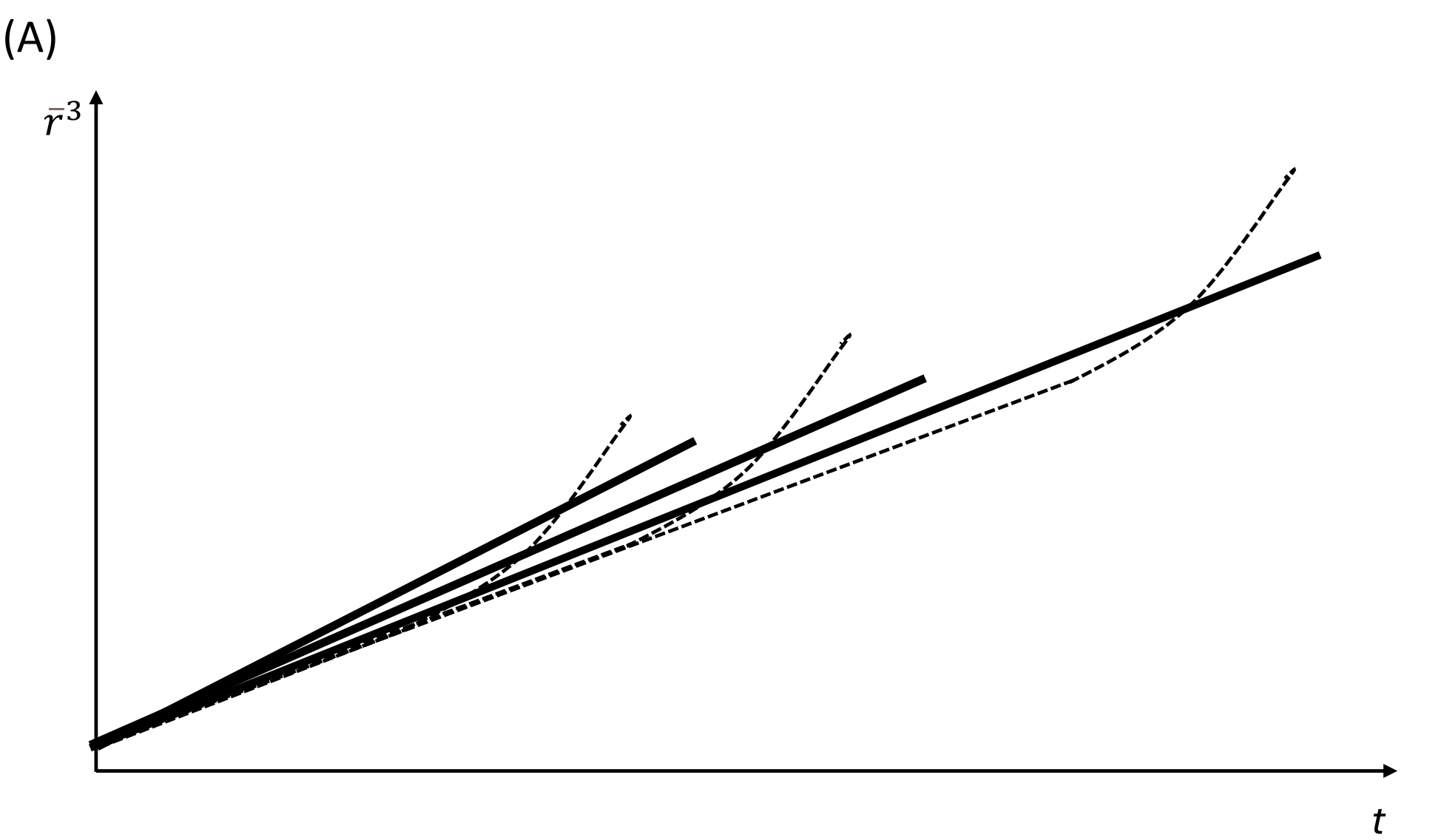

(B)

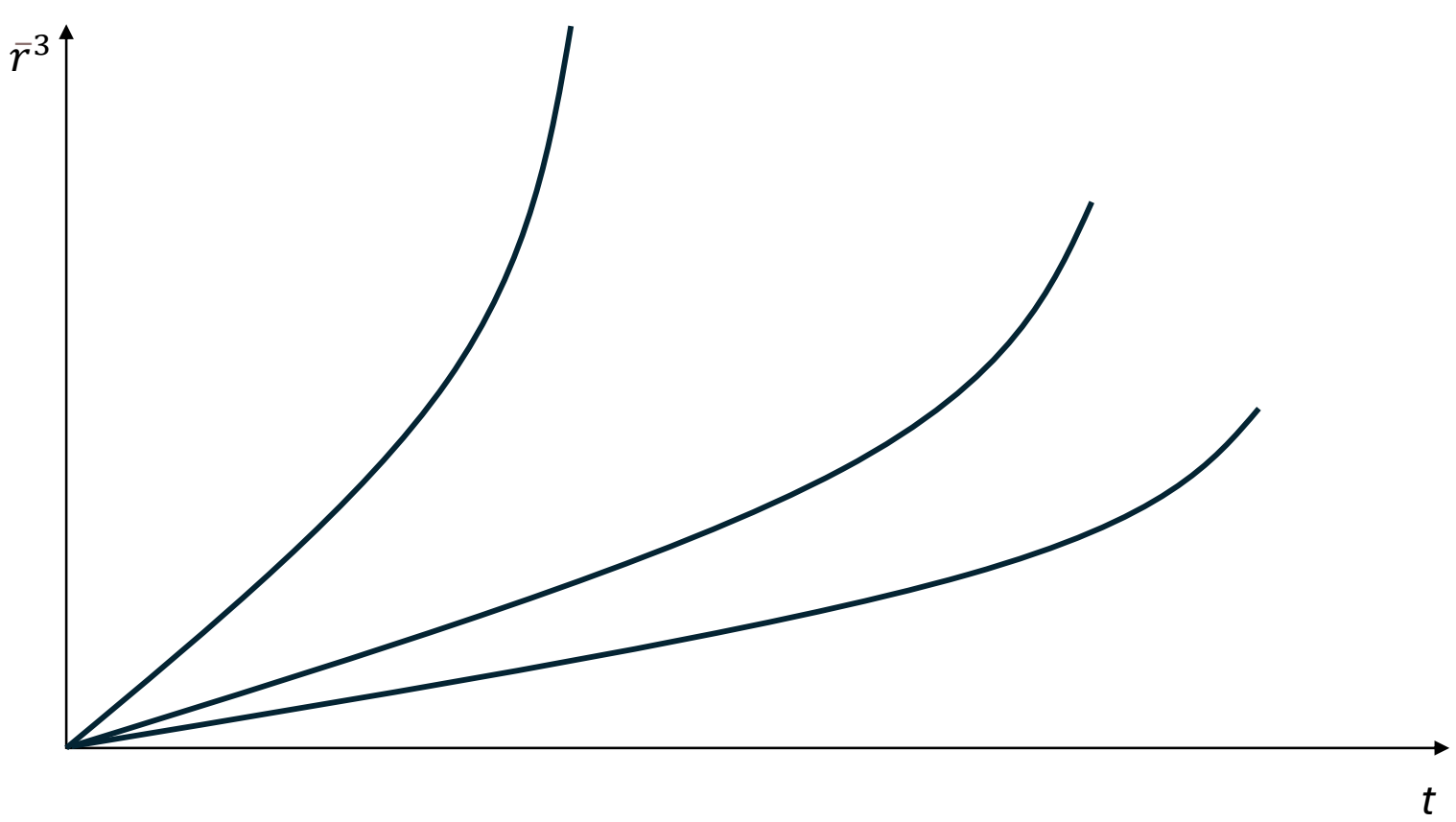

(C)

Figure 2. Cartoon illustrating the transition from the cubic law of the average particle size growth to quadratic law as a function of the concentration atmosphere parameter $\kappa$. (A) On the graph, $1/\kappa_3 > 1/\kappa_2 > 1/\kappa_1$, which corresponds to the increase in the micellar volume fraction in this order. (B) If the non-linearity of the plots is ignored and the dependencies are forced into cubic coordinates, increasing the micellar volume fraction is predicted to increase the cubic ripening rate.(C) If the range of particle sizes is such that $r \gg 1/\kappa$, the initial linear part of the graph may not be visible and the plots will look concave, with increasing curvature at higher micellar concentrations.

The paper will have the following outline. In the next section, we will review the contributing factors to the micelle solubilization source-sink term $\omega$, which will enable to determine the points of transition from the cubic to kinetic growth as a function of the micellar concentration and provide some estimates of micellar effects on the ripening rate in emulsions. In the next section, we compare the proposed model with the experimental data generated in series of studies of Weiss et al [8, 9], including the linearity aspects, and conduct the comparison the absolute rates with the theory.
The paper also has three Appendixes. In the Appendix 1, we re-derive eqn (7), which is very similar to eqn 34 of Ref. [4]; the main difference is that the micelles in the ripening case are (over)saturated with the solubilizate, whereas in the solubilization kinetics experiment, our primary interest was the case of 'empty' micelles, with no solubilizate in them. In Appendix 2, the effect of Ostwald ripening-related supersaturation on micellar solubilization is discussed, and it is shown that the adjustment to micellar structure is expected to be relatively minor; this has been a controversial point in literature [19]. Finally, in Appendix 3, we discuss the parameter selection for the estimates of the absolute ripening rates, specifically, the aqueous solubility and diffusivity of hydrocarbons, micellar size, and assumed value of the energy potential barrier for SDS micelles; we also discuss the possible reasons for the disagreement between the experimentally determined values of the ripening rates and the theoretical estimates.

**Micelle source-sink term and its effect on ripening kinetics**

Consider Ostwald ripening in O/W presence of micelles. In dilute-in particles regime, the micelles are in equilibrium with the aqueous solution of the oil in water, $C_m(t)$ in infinity. As the oil is in excess in the system and is emulsified, the micelles are saturated with the solubilizate; in fact they are slightly over-saturated over the level of the macrophase ($r = \infty$) due to the excess chemical potential of oil in emulsion drops, which is due to the surface tension and Laplace pressure effects; see Appendix 2 for more details on how does the micelles respond to such a supersaturation. If a micelle is located by an emulsion drop, it will either absorb more oil or release some oil, depending on if the local concentration $C$ where the micelle is located is larger or smaller than $C_m(t)$ . This will produce an effective source/sink for the oil in the given location. Assuming that there is no energy barrier for the oil molecule, the flux of the oil into a spherical micelle is equal to:

$$j = 4\pi r_{mic} D(C - C_m) = 4\pi r_{mic} D\Delta C \equiv \omega_0 \Delta C \qquad (10)$$

The equation is essentially the diffusional flux of the oil out the sphere of the micellar size; the flux is positive if the micelle is locally oversaturated with the solubilizate and negative in the opposite case. We use dimensionless concentration units, which is the weight concentration reduced to the solute density; accordingly, the unites for $\omega_0$ are $cm^3/s$; $r_{mic}$ is the micellar radius. The equation assumes that there is no barrier for the oil to get from the aqueous solution to the micellar interior and backwards and the process is diffusion controlled. However, there could be such a barrier in existence [20,21]; to account for it we could add an Arrhenius factor to the equation:

$$\omega = \omega_0 exp\left(-\frac{\Delta G}{RT}\right) \qquad (11)$$

where $\Delta G$ is the activation free energy for the oil to enter or leave the micelle. In that sense, $\omega_0$ can be considered the pre-exponent and the oil exchange cannot proceed faster than this rate; then the source-sink term becomes equal to:

$$\omega = 4\pi r_{mic} D\ exp\left(-\frac{\Delta G}{RT}\right) \qquad (12)$$

In our previous paper of this series [21], we analyzed the energy barrier required for a hydrocarbon molecule to enter a micellar interior. According to this model, hydrocarbons can freely enter nonionic micelles of ethylene oxide type, regardless of their molecular weight. However, this is not the case for ionic micelles; as a hydrocarbon molecule approaches an ionic micelle, it is repelled by electrostatic forces because of the low molecular polarizability of hydrocarbons; the repulsion per molecule is predicted to increase with increasing the molecular weight of the hydrocarbons. As the value of the energy barrier changes with the separation distance, it is convenient to operate not in terms of the energy barrier as such, but in terms of the net reduction factor for the sink-source term:

$$F^* = \frac{\omega_0}{\omega} \qquad (13)$$

Here $F^*$ can be interpreted as a reciprocal of the Arrhenius factor averaged over the separation distance, and

$$\omega = \frac{4\pi r_{mic} D}{F^*} \qquad (14)$$

Finally, eqn (14) the source/sink term for a single micelle; if the number of micelles per unit volume is $n_{mic}$, the total source-sink strength becomes:

$$\omega n_{mic} = \frac{4\pi r_{mic} n_{mic} D}{F^*} \qquad (15)$$

or

$$\omega n_{mic} = \frac{3\phi_{mic} D}{F^* r_{mic}^2} \qquad (16)$$

where $\phi_{mic}$ is the micellar volume fraction; the micelles are assumed to be spherical with the volume equal to $\frac{4\pi r_{mic}^3}{3}$. Accordingly,

$$\kappa = \frac{1}{r_{mic}} \sqrt{\frac{3\phi_{mic}}{F^*}} \qquad (17)$$

In our previous study [21], we calculated the values of $F^*$ and $\kappa$ for micelles of sodium dodecyl sulfate (SDS) and Tween 20; the values are shown in Table 1. The table shows the approximate values for n-decane, n-tetradecane, and n-hexadecane at various values of the micellar volume fractions. In calculations, the micellar size and shape were assumed to be constant and independent on the surfactant concentration. We note that the values of $1/\kappa$ are smaller for Tween 20 than those for SDS, as the micelles of ethylene oxide type can absorb oils without a potential barrier. For SDS, there is a potential barrier in place, and it is predicted to increase with the hydrocarbon chain length. The origin of the effect is argued to be the exclusion of the hydrocarbon molecules from the double layer area around the ionic micelles, where there is an electrostatic energy penalty for hydrocarbon molecules to enter into the area of high electrical field strength. Naturally, the energy penalty per molecule increases with the molecular volume of hydrocarbons. In our previous study [21], we evaluated this energy penalty and compared it to the values extracted from the experimental rates of the solubilization kinetics, and a fair agreement was obtained. These two factors, that is, the micellar size and surface charge have a direct influence on the micellar effects on the Ostwald ripening kinetics.

Table 1. Values of $1/\kappa$ for the mass transfer of n-decane, n-tetradecane, and n-hexadecane in micellar solutions of sodium dodecyl sulfate (SDS) and Tween 20, at different volume fractions of micelles*.

| $\phi_{mic}$ | 0.0001 | 0.001 | 0.0034 | 0.01 | 0.1 |
|---|---|---|---|---|---|
| $1/\kappa$, nm, Tween 20, decane, tetradecane and hexadecane | 185 | 58.4 | 31.7 | 18.5 | 5.84 |
| $1/\kappa$, nm, SDS, decane | 492 | 156 | 84.6 | 49.2 | 15.6 |
| $1/\kappa$, nm, SDS, tetradecane | 803 | 254 | 138 | 80 | 25 |
| $1/\kappa$, nm, SDS, hexadecane | 1270 | 402 | 218 | 127 | 40 |

*Parameter set used in calculations is discussed in Appendix 3.

## Comparison to experimental data

Ostwald ripening is an important factor in stability of submicron emulsions used in pharmaceutical and food industry [18]. The experimental studies of Ostwald ripening in emulsions in presence of micelles are now quite extensive, see Refs [5-11]. For ionic surfactants, particularly SDS, it is in general agreed that the effects of micelles on the ripening rate are moderate, and the ripening rates are only (at most) factor of 2-10 larger than those observed in the absence of micelles, despite the significant solubilization capacity of the micelles [5,6,7]. To validate the predictions of this paper, the study of ripening over a wider range of particle sizes is required. The study of Weiss et al [8] was conducted over very wide range of particle sizes, from 0.05 to 1 μm, and utilized two different particle sizing techniques, static light scattering and ultrasound scattering. The wide range of particle sizes places it apart from the rest and it would be interesting to discuss its results versus the prediction of this work. The graphs in Figure 3 show the dependence of the radius cubed on time for decane and hexadecane emulsions stabilized by 20 mM SDS. The decane graph does show a significant concavity over all the range. On the other hand, the hexadecane graph shows less of concavity. The prediction of our study is that the first graph should become concave after reaching the particle radius of about 0.085 μm; whereas the second graph should become concave at about 0.218 μm. Both these predictions are difficult to validate; for the first graph, the transition point is squeezed between the few first points; and for the second it is outside the range of the plot, which means the graph as shown must be mostly cubic. Yet, the fact that the first graph is more concave than the second seems to support the model. In order to better validate the predictions of this paper, the experimental data should cover both the small and large average particle size ranges. If more data points would have existed over the initial stage of the kinetics for decane over small particle size range, this prediction could have been better validated.

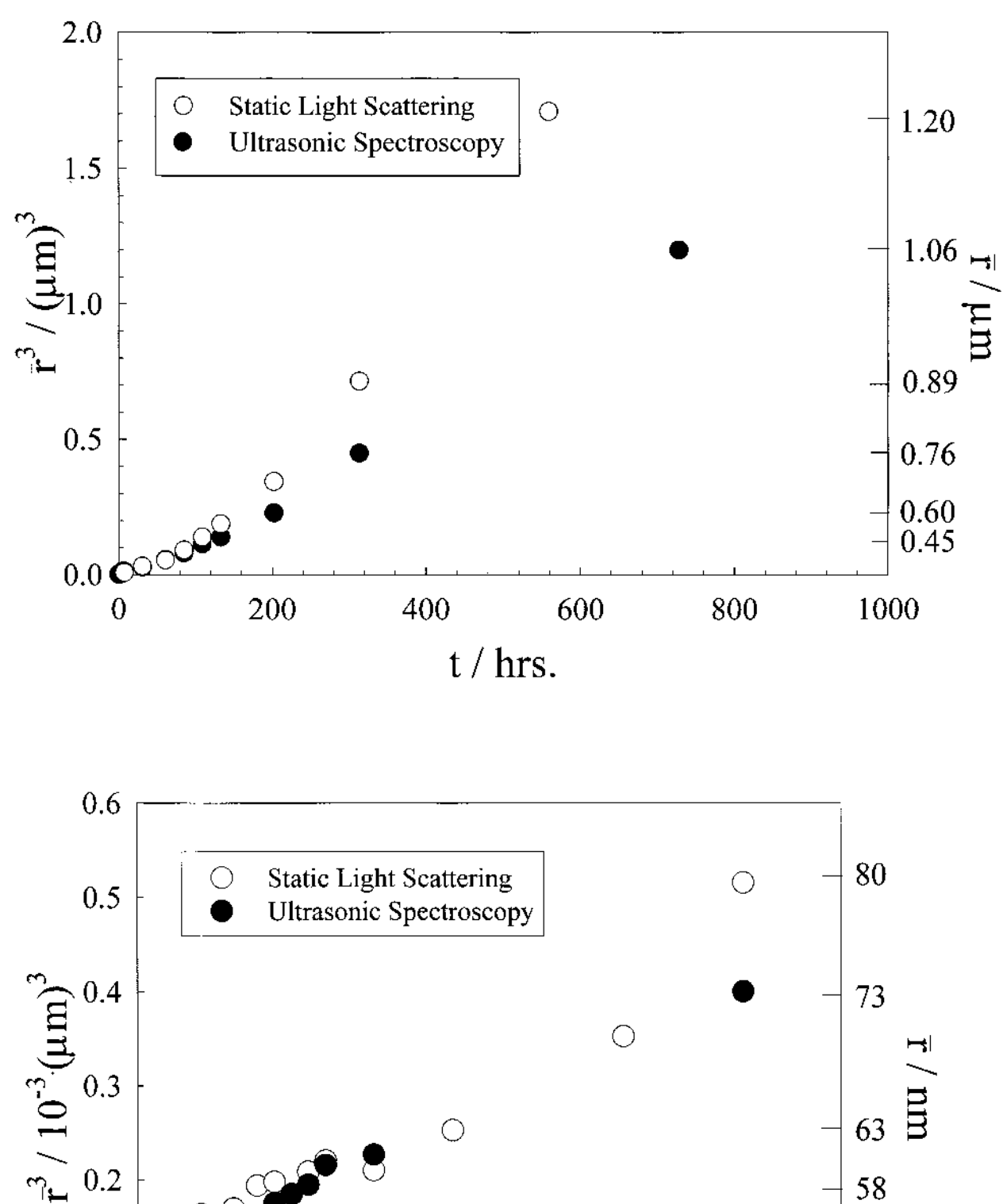

Figure 3. Ostwald ripening of decane (top) and hexadecane (bottom) emulsions stabilized by 20 mM SDS. Reproduced, with permission, from Ref 8. Copyright of ACS.

The data of the plots were re-cast in quadratic coordinates and their linearity was evaluated in Figure 4 and Table 2. One can confirm that indeed the quadratic fit is slightly better for decane and slightly worse for hexadecane for both particle sizing techniques.

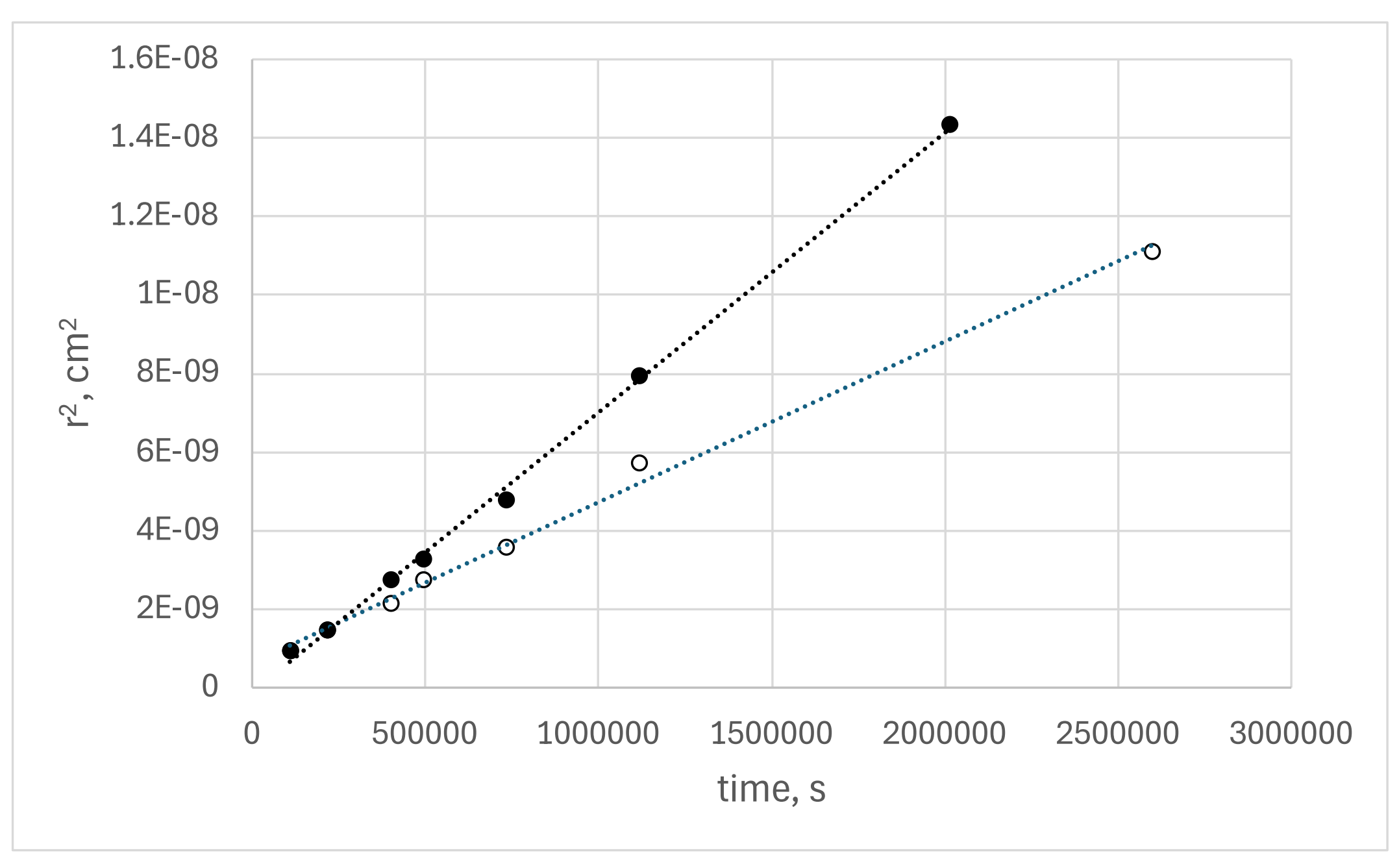

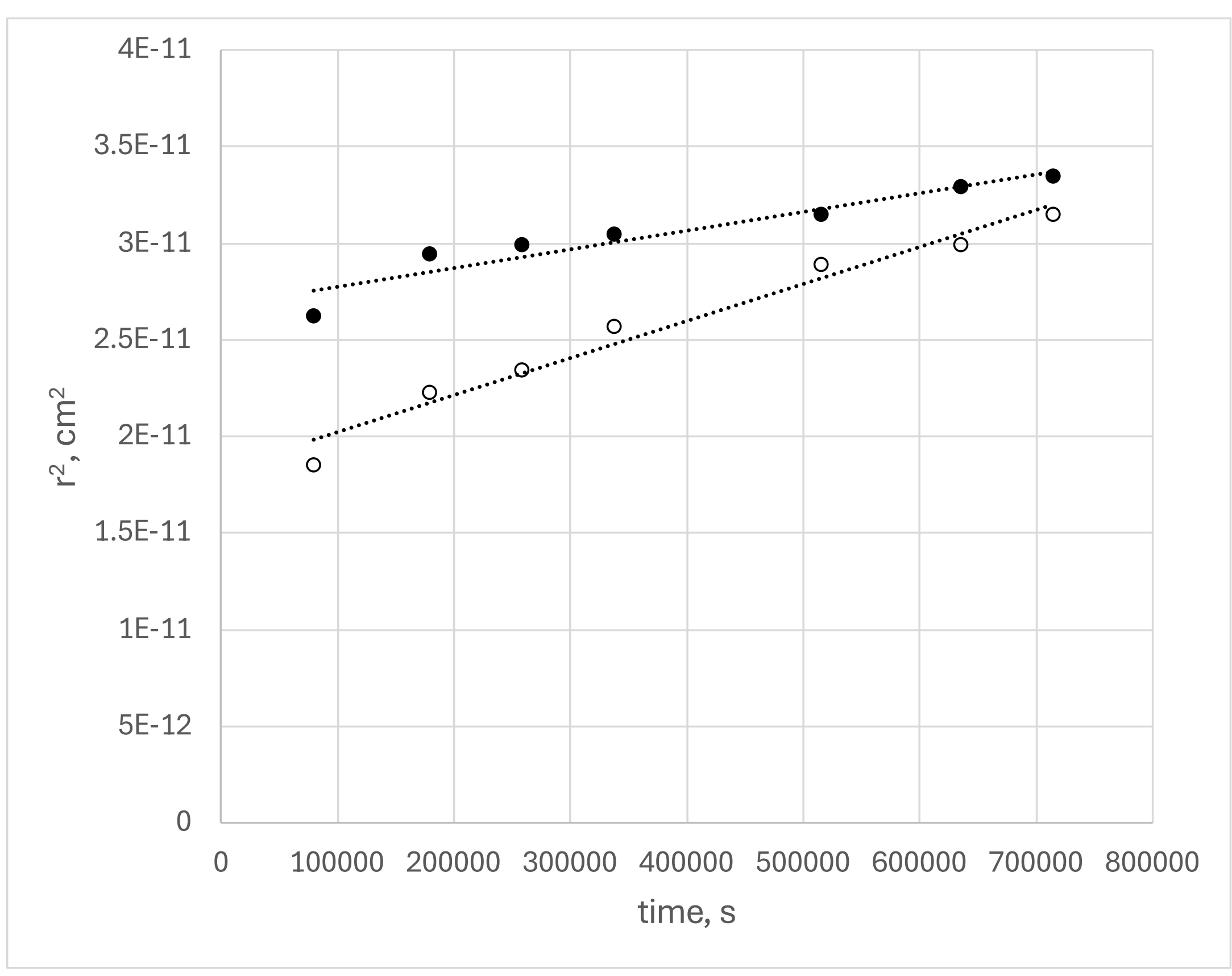

Figure 4. Data of Figure 3 re-cast in quadratic coordinates. The plots show slightly improved linearity for decane (top), but not for hexadecane (bottom), agreeing with the arguments of this paper.

Table 2. Linearity of quadratic and cubic fits for the systems of Figures 3 and 4.

| oil | quadratic correlation coefficient, SLS | quadratic correlation coefficient, ultrasound | cubic correlation coefficient, SLS | cubic correlation coefficient, ultrasound |
|---|---|---|---|---|
| decane | 0.9984 | 0.9954 | 0.9721 | 0.9923 |
| hexadecane | 0.9049 | 0.9698 | 0.915 | 0.9801 |

It would be also of interest to compare the absolute values of the quadratic rates for Ostwald ripening of decane emulsions with the prediction of eqn (9), which is made in Table 3. The theoretical rate for decane is quite close to the experimental values, differing only by the factor of 1.5 - 2.5, depending on the particle sizing method. The difference can be explained by many factors, one among others being the difference in particle size averaging techniques employed by static light scattering and ultrasound scattering. Ripening theories [1-3] operate in terms of number average radius, whereas both scattering methods tend to produce the values closer to the weight average radius, which is expected to be larger than the number-averaged radius. More reasons for the possible differences are described in Appendix 3.

Table 3. Quadratic Ostwald ripening rates, $v$, $cm^2/s$, of decane emulsions stabilized by SDS extracted from Figure 4, compared to the theory of this paper. See Appendix 3 for the selection of the parameters for the theoretical calculation; $1/\kappa$ = 84 nm. The ratio of the experimental rate to the theoretical one is shown in parentheses.

| quadratic rate, ultrasound scattering | quadratic rate, SLS | theory, eqn (9) |
|---|---|---|
| $4.0X10^{-15}$ (X1.40) | $7.3x10^{-15}$ (X2.56) | $2.85x10^{-15}$ |

For hexadecane, on the other hand, within the particle size range studied, the cubic fit should be more appropriate; the comparison of absolute rates is made in Table 4. The agreement there is not as good, as the system is expected to transition from the cubic to quadratic mode. The difference between the theory and experiment here is about a factor of 7. Note that even bigger difference, a factor of 20, is seen when the data are subjected to quadratic fit.

Table 4. Cubic Ostwald ripening rates, $w$, $cm^3/s$, of hexadecane emulsions stabilized by SDS extracted from Figure 4. See Appendix 3 for the selection of the parameters for the theoretical calculation. $1/\kappa$ = 218 nm. The ratio of the experimental rate to the theoretical one is shown in parentheses.

| cubic rate, ultrasound scattering | cubic rate, SLS | theory, eqn (5) |
|---|---|---|
| $1.0X10^{-22}$ (X6.7) | $1.1 x10^{-22}$ (X7.3) | $0.15x10^{-22}$ |

We are going to discuss now the ripening rates of nonionic surfactants, which are in general faster than those for ionic surfactants [8,9]. The effect is significant and cannot be attributed to the surface tension differences [8]. This has been shown, in particular, in the study of Ostwald ripening of tetradecane and a variety of nonionic surfactants of ethylene oxide type [9]. Fortuitously, the authors even fitted the data in quadratic coordinates, as they were validating their model of Ostwald ripening enhancement by surfactant micelles based on collisions between the micelles and emulsion drops [22]. The authors concluded that the quadratic fit for nonionic systems works better than the cubic fit. They also studied the dependence of the quadratic ripening rate on the surfactant concentration; in most cases, the dependence was slower than linear. In some cases, they even observed the rate to decrease at very high surfactant concentrations, see Figure 5.
Analyzing this data within the framework of this paper, for the surfactant concentration and particle size ranges studied, the systems were well within the range of $1/\kappa \ll r$, and, and the data should follow the quadratic (Wagner) limit, with the rate of:

$$v = \frac{d(\bar{r}^2)}{dt} = \frac{64 C_0 \sigma V_m D \kappa}{81RT} = \frac{64 C_0 \sigma D V_m}{81RT} \frac{\sqrt{\frac{3\phi_{mic}}{F^*}}}{r_{mic}} \qquad (18)$$

Note that for nonionic surfactants such as Tween 20, the barrier function $F^*$ is assumed to be 1. If the quadratic ripening rate is plotted as a function of the square root of micellar volume fraction, the graph should be linear, with the slope *s* equal to $s=\frac{64\sqrt{3}C_0\sigma D V_m}{81 r_{mic} RT}$, provided that the micelles conserve their size and shape as the surfactant concentration increases. The latter is not always the case, as in many cases micellar growth and the transition from spherical to wormlike micelles occurs with increasing the surfactant concentration. Figure 6 shows the dependence of the ripening rate on the squared root of the micellar volume fraction. If the slope is measured as based on the first four points, the value of the slope comes to $12x10^{-16}$ $cm^2/s$, which is quite close the theoretical value as calculated by eqn (18), $3.78x10^{-16}$ $cm^2/s$, but 3.2X larger. The absolute ripening rates calculated by eqn (18) are also in a good agreement with the experiment, as shown in Table 5; again, the experimental rates are larger than theoretical and the difference, at least partially, can be attributed to the difference of the particle size averaging method from the number-based averaging. We attribute the decrease in the rate at higher surfactant concentrations to the micellar shape transition. More possible reasons for the differences are discussed in Appendix 3.

Table 5. Comparison of quadratic Ostwald ripening rates for tetradecane-in water-emulsions stabilized with Tween 20, with theory. For the discussion of the parameters of the model, see Appendix 3. The ratio of the experimental rates to the theoretical ones is shown in parentheses.

| $c$, wt% | $\phi_{mic}$ | $v$, $cm^2/s$, experiment | theory, eqn (18) |
| --- | --- | --- | --- |
| 0.5 | 0.0045 | $6.7 X 10^{-17}$ (X3.35) | $2.0 X 10^{-17}$ |
| 1 | 0.0091 | $1.2 X 10^{-16}$ (X4.3) | $2.8 X 10^{-17}$ |
| 2 | 0.0182 | $1.5 X 10^{-16}$ (X3.75) | $4.0 X 10^{-17}$ |
| 5 | 0.0455 | $2.6 X 10^{-16}$(X4.1) | $6.3 X 10^{-17}$ |

Note that equation (18) has no adjustable parameters and with the view that the solubility of tetradecane in water is not known very precisely, this level of agreement is remarkable. The possible reasons explaining the existing differences are discussed in Appendix 3.

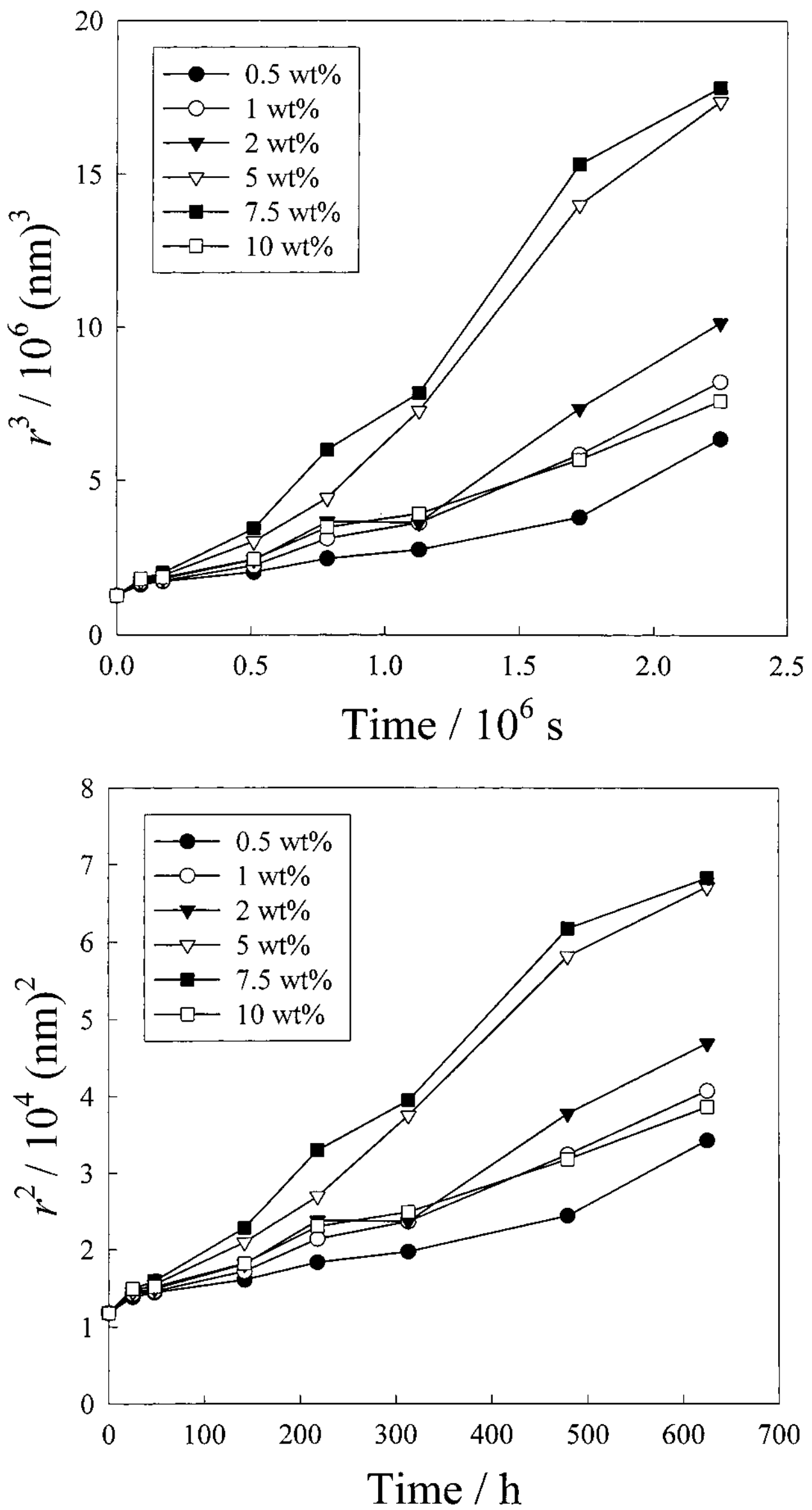

Figure 5. Ostwald ripening of tetradecane-in water emulsions stabilized by Tween 20, in cubic and quadratic coordinates. Reproduced, with permission, from Ref [9]. Copyright of ACS.

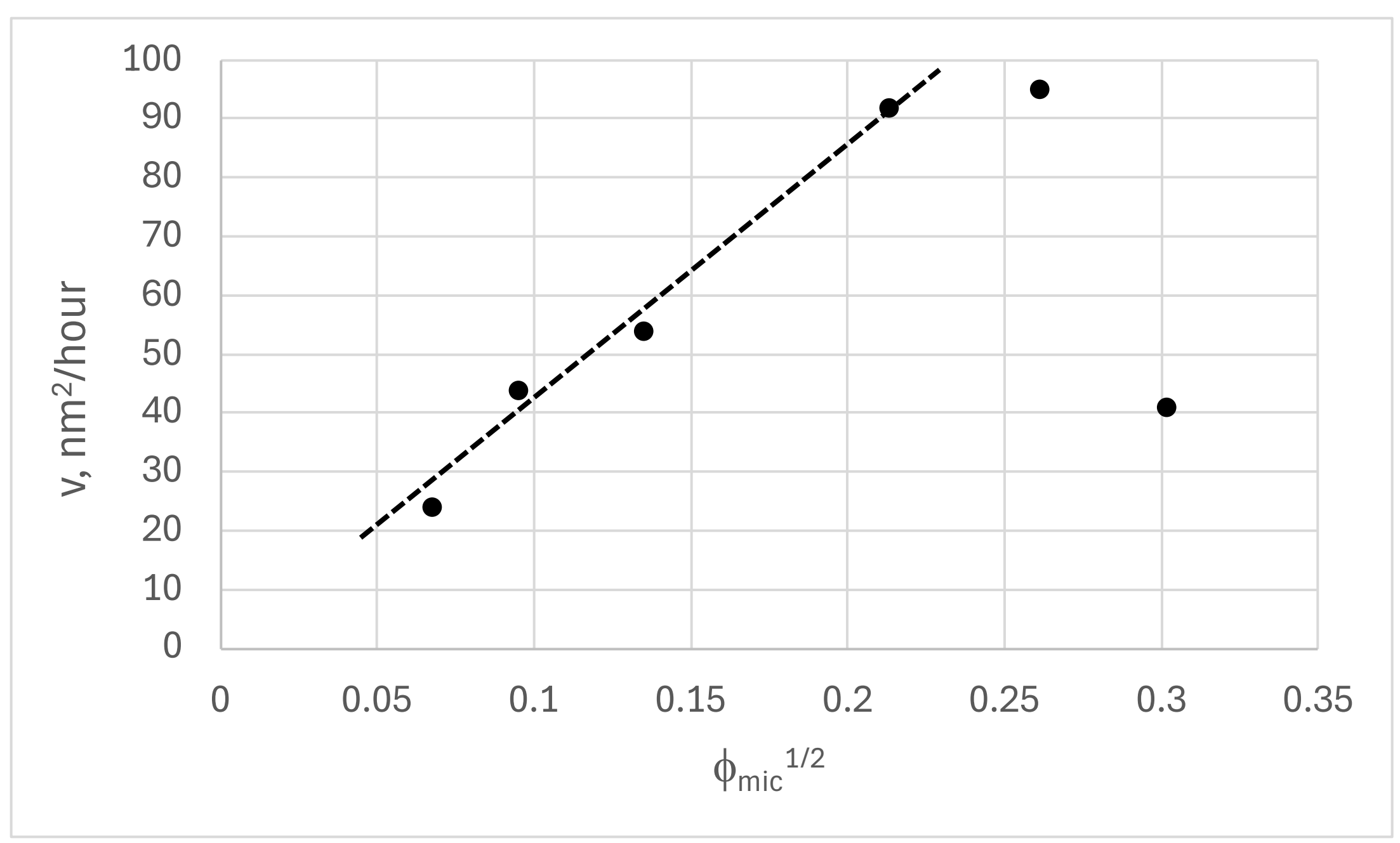

Figure 6. Quadratic Ostwald ripening rates of tetradecane-in-water emulsions stabilized by Tween 20 plotted versus square root of micellar concentration. The data are calculated based on Table 1 of Ref. [9].

**Conclusions**

In this paper we theoretically studied Ostwald ripening in O/W emulsions in presence of solubilizing micelles. We explored the dilute-in-emulsion droplet case, with the particles separated from each other by distances much larger than their sizes. The prediction of the theory is that, at very small average sizes, the particle size growth follows the classical LSW cubic law. In this regime, eqn (5) in which $C_0$ is the molecular solubility of the oil in water, still fully holds. The kinetics transitions to the quadratic law at larger particle sizes, that is, to eqn (9), where $\kappa$ is determined by eqn (17). The crossover point of the kinetics depends on the dynamics of the micellar exchange; it is set by the value of the oil atmosphere distribution parameter, $\kappa$, which, somewhat like Debye length, is proportional to the square root of the, in this case, micellar concentration. It should be noted that in the range when $1/\kappa$ is close to $r$, the ripening kinetics still can be forced into the cubic law, with only moderate deviations; in this case, experimentally, the micellar effects are seen not so much as the deviations from the linearity, but as an apparent increase in the cubic rate. The predicted transition from the cubic to quadratic law still need to be better validated, as the available experimental data do not sufficiently support it. A study over wider range of average sizes, covering both the small and large particle size ranges would have enabled it.
The predicted in this paper linear increase of the squared average radius with time in presence of micelles should not be confused with other processes that can follow the same kinetics, be it diffusion through a poorly penetrable membrane, or ripening by direct attachment of micelles to the emulsion drops[22,23]. Although the linear increase of the squared radius with time is expected there too, this is where the similarity ends; the mechanisms of the processes are quite different. The advantage of the model of this paper is that it provides the quadratic ripening rate in an explicit form, without any empirical parameters. It also explains why the ripening rate correlates with the aqueous solubility of hydrocarbons so well, which the model based on collisions between the micelles and emulsion drops has difficulty to explain.
Finally, the model has been developed for dilute emulsion systems in presence of micelles. There has been a significant interest in studies of Ostwald ripening in concentrated systems, where the particles separated by distances comparable or smaller than their sizes [24]. In those systems, a moderate increase in the rate with the volume fraction of the dispersed phase is predicted. The effect is due to shortening of the diffusion path as the volume fraction of the particles increases. We would like to emphasize that in micellar systems, this effect can be seen only if $r<<1/\kappa$. In the opposite limiting case, $r>>1/\kappa$, no such increase in the rate is expected and the ripening is predicted to be largely independent on the volume fraction of the drops, as the diffusion path $1/\kappa$ is controlled by the micelles. More studies are required for this case to analyze and validate it.

**Appendix 1. Derivation of mass transfer Eqn (7) for Ostwald ripening case**

The derivation is analogous to the one proposed in Ref. [4] for the solubilization kinetics. Consider the mass transfer from an emulsion drop undergoing ripening; the drop is in a micellar solution. We are putting down the continuity equation for the oil concentration surrounding the particle:

$$\frac{\partial C}{\partial t} = -D\nabla^2(\Delta C) + \omega n_{mic}\Delta C \qquad (19)$$

Here $\Delta C = C(\rho, t) - C_m(t)$ is the concentration difference over the one at the infinity; $\omega n_{mic}\Delta C$ is the micellar source term as discussed above. Micelles are in equilibrium with the medium at the infinity and absorb or release the oil depending on if $C(\rho, t)$ is bigger or smaller than $Cm(t)$ locally. We use the letter $\rho$ to indicate the distance from the center of the particle in spherical coordinates, to be distinguished from the value of the particle radius, $r$, which is constant. We will be looking for the steady state of the diffusion process. Equating $\frac{\partial C}{\partial t} = 0$, we are getting:

$$-D\nabla^2(\Delta C) + \omega n_{mic}\Delta C = 0 \qquad (20)$$

The solution for this equation in spherical symmetry case is like the one shown in Ref [4]:

$$\Delta C(\rho) = C(\rho) - C_m = (C_p - C_m)\frac{r exp(-\kappa\rho)}{\rho exp(-\kappa r)} \qquad (21)$$

Here $C_p$ is the concentration at the surface of the particle, at $\rho = r$, and $\kappa$ is determined by eqn (7). Equation (20) is analogous to Poisson equation of the electric double layer; $1/\kappa$ is somewhat akin to the Debye length; just like the Debye length, it is reciprocal in the square root of, in this case, micellar, concentration. The parameter $\kappa^{-1}$ plays a similar role here to the screening length of Poisson-Boltzmann equation of electrostatics, with the difference that it determines not the distribution of electrical charge but that of the molecular solution of the diffusing dissolved oil.
The material flux out of the particle can be obtained by the differentiation of eqn (21):

$$J = 4\pi r^2 D\nabla C|_{\rho=r} \qquad (22)$$

and eqn (7) is recovered.
Brownian motion of the micelles is another factor that complicates the matter. The conceptual flow above assumes that the micelles are not mobile as they participate in the oil exchange with the medium. However, they cannot finish the mass transfer as they are replaced by other 'fresh' micelles that are coming in constantly. The assumption is that it truly does not matter as some partial mass exchange does happen and some fresh micelles are always in place to participate in the mass exchange and the incoming mass flux becomes distributed over several micelles instead of one.

**Appendix 2. Estimate of the micelle swelling due to the excess oil chemical potential of oil during Ostwald ripening of O/W emulsions.**

Consider a spherical micelle in equilibrium with excess solubilisate. We will be using simplified Safran et al approach to the problem [25], assuming the free energy *G* to driven by the bending free energy of the monolayer, and ignoring the saddle splay modulus:

$$G = 4\pi r_{mic}^2 \kappa_b \left(\frac{1}{r_{mic}} - H_0\right)^2 \quad (23)$$

where $H_0$ is the surfactant monolayer spontaneous curvature, $\kappa_b$ is the bending modulus of the monolayer. The radius of the spherical micelle at the 'solubilizaton failure line' is controlled by the ratio of the volume of the micelle to its surface area:

$$r_{mic} = \frac{3(n^{oil} v_m^{oil} + n^{surf} v_m^{surf})}{n^{surf} A^{surf}} \quad (24)$$

here $n^{oil}$ and $n^{surf}$ are the numbers of oil and surfactant molecules in the micelles, $v_m^{oil}$ and $v_m^{surf}$ are the volumes of the surfactant and oil molecules, and $A^{surf}$ is the area per surfactant molecule in the surfactant monolayer.

The free energy has a minimum when the micellar radius is equal to the reciprocal spontaneous curvature, $r_{mic} = 1/H_0$; if more oil is added into the micelle, the free energy increases. The chemical potential of the oil in the micelle per molecule, in the vicinity of the saturation (aka emulsification failure) point is equal to:

$$\Delta\mu_{oil} = \frac{\partial G}{\partial n_{oil}} = \frac{\partial G}{\partial r_{mic}} \cdot \frac{\partial r_{mic}}{\partial n_{oil}} = \frac{8\pi\kappa_b}{n_{oil}}(1 - H_0 r_{mic})\left(1 - \frac{v_m^{surf} H_0}{A_{surf}}\right) \quad (25)$$

The chemical potential reference point is set at zero for the macrophase when the micellar size is equal to $1/H_0$; for emulsion drops there is an excess value of the chemical potential over this value by the product of the Laplace pressure and molar volume of oil:

$$\Delta\mu_{oil} = \frac{2\sigma V_m}{r} \quad (26)$$

For 0.1 μm drops, the increment is about 0.01 $k_B T$ per molecule. Assuming the monolayer bending modulus of 1 $k_B T$, the number of oil molecule solubilizates per micelles of 10, and ignoring the $\frac{v_m^{surf} H_0}{A_{surf}}$ term as a small correction, we conclude that the radius of the micelle deviates from the equilibrium value by less than percent in the response to such supersaturation, which the micelles are expected to easily accommodate.

**Appendix 3 Parameters used in calculations**

Theoretical calculations of ripening rates require the knowledge of aqueous solubility of hydrocarbons, their diffusivities, interfacial tensions, and molar volumes. The most uncertain part in this calculation is the value of the aqueous solubility of hydrocarbons. This work utilized the data of McAuliffe [26], [27] that extend to decane, with the following extrapolation to tetradecane and hexadecane assuming exponential decrease in solubility with the chain length (i.e., the same value of the free energy penalty per each $CH_2$ group). The validity of this approach has been recently confirmed by the theoretical conformational studies of alkyl chains in water [28]. The diffusion coefficients were calculated by Hayduk-Laudie equation [29], with the assumption of the viscosity of the solution being equal to 1 cPoise; the values are similar to the ones calculated by Wilke-Chang equation [30]. The experimental values of the interfacial tensions from literature were used. The data are summarized in Table 6.
The calculation of $\kappa$ requires the knowledge of the micellar size and the energy barrier for hydrocarbons to enter a micelle. The micellar radius was assumed to be equal to 2.2 and 3.2 nm for SDS and Tween 20, respectively [31,32], and the micellar polydispersity was ignored. The micellar size is expected to be larger than this due to the micellar size increase because of the solubilization; this aspect is neglected as well. The energy barrier factor $F^*$ was assumed to be equal to 15, 40 and 100 for decane, tetradecane, and hexadecane, respectively. For Tween 20, the value of $F^*$ was assumed to be equal to 1 for all hydrocarbons. These values are based partially on solubilization kinetics data and partially on theoretical estimates, as described in [21]. The micellar volume fraction was calculated based on concentration of surfactant and its density, which was 1.0 and 1.1 g/cm$^3$ for SDS and Tween-20, respectively. The depletion of the surfactant from the bulk because of its adsorption at the oil-water interface was neglected.

Table 6. Physical properties of hydrocarbons in surfactant solutions

| Hydrocarbon | density-reduced solubility, ml/ml Ref [5]; extrapolation is based on [19], [20], [21] | Diffusivity, cm$^2$/s Hayduk-Laudie equation | Molar volume, cm$^3$/mol | interfacial tension vs SDS or Tween 20, dyn/cm |
|---|---|---|---|---|
| n-decane (vs. SDS) | $7.1 \times 10^{-8}$ | $5.94 \times 10^{-6}$ | 195 | 9.0 [5] |
| n-tetradecane (vs. Tween-20) | $3.7 \times 10^{-10}$ | $5.01 \times 10^{-6}$ | 260 | 3.5 [8] |
| n-hexadecane (vs. SDS) | $2.7 \times 10^{-11}$ | $4.66 \times 10^{-6}$ | 294 | 10.8 [5] |

An important step in validation of the model was the comparison of the experimental absolute ripening rates, both quadratic and cubic, with the theory. The level of agreement for quadratic rates was between the factors of 2 and 4. The cubic rate value for hexadecane is overestimated more, by X7. Note that there were no adjustable parameters used in the fit and all the parameters were determined independently from other sources. Note that *F** is not an adjustable parameter; its value comes from the electrostatic calculations for SDS micelles and validated against the solubilization kinetic data for SDS-hydrocarbons per Ref [21]. The level of agreement between the absolute rates of ripening with the theory requires further discussion. Below we describe the possible reasons for these differences.

1. Systematic experimental error in particle sizing (i.e., the difference between the average radius as measured and the number averaged value). The scattering particle sizing methods involve an inverse transformation process that weigh different fractions as based on their scattering intensity; thus, in Rayleigh limit of scattering, it generates a weight-average value. Mie range with the particle sizes comparable to the wavelength of light produces neither weight- nor number average, and the correction factor may depend on the particle size. Note that a moderate systematic error in the average particle size of 33% can lead to 66% error in quadratic rate of ripening and 2X error in cubic rate. The observed differences in ripening rates as based on the static light scattering and the ultrasound scattering provide an idea of the expected magnitude of this effect.
2. The system has not reached the asymptotic regime. LSW theory strictly applies at the very late stage of ripening, when the distribution function reached the attractor stage. This condition may not have been fully met in experiment.
3. The parameters of the models are not known exactly. The largest error here comes from not knowing the aqueous solubility of hydrocarbons and is based on an extrapolation as described above in this section. The diffusion coefficients were calculated/extrapolated as well but the error there is expected to be smaller. For the ionic surfactants, there could be more uncertainties. The value of the energy barrier *F** is not known precisely; the experimental values of *F** determined from the solubilization kinetics also show some variability [21].
4. The model also has intrinsic simplifications, such as the assumption of micellar mono-dispersity and the independence of the micellar size and shape on the surfactant concentration. Although the latter can be the case for some systems, it is not the case for some others where micellar growth and transition to non-spherical (wormlike) micelles is observed with increasing the surfactant concentration [33]. On top of this, there are also simplifications that the model cannot properly address; for example, the Brownian motion of emulsion drops and micelles are not accounted for; both are expected to increase the experimental ripening rate. In our previous study of Ostwald ripening of hydrocarbon emulsions in cubic regime, Ref. [5], even after attempts to correct for the particle averaging differences, there was still a factor of 2.5X difference remaining that was attributed to the Brownian motion of emulsion drops; it also can be the case for the quadratic rates of this study.